\documentclass[12pt,preprint]{aastex}

\def\spose#1{\hbox to 0pt{#1\hss}} 
\def\simlt{\mathrel{\spose{\lower 3pt\hbox{$\mathchar"218$}}      
\raise 2.0pt\hbox{$\mathchar"13C$}}} 
\def\simgt{\mathrel{\spose{\lower 3pt\hbox{$\mathchar"218$}}      
\raise 2.0pt\hbox{$\mathchar"13E$}}} 
  
\def\etal{{\rm et~al.~}} 

\revised{Sept 3, 2008} 
\lefthead{Mould \& Sakai} 
\righthead{Extragalactic Distance Scale } 
\begin{document}
\title{The Extragalactic Distance Scale without Cepheids}

\author{Jeremy Mould} 
\affil{School  of Physics, University of Melbourne, Vic 3010, Australia} 
\authoraddr{E-mail: jmould@ph.unimelb.edu.au} 
\author{Shoko Sakai}
\affil{Dept of Physics \& Astronomy, UCLA, Los Angeles, CA 90095}
\authoraddr{E-mail: shoko@astro.ucla.edu} 


\begin{abstract}
Distances of galaxies in the Hubble Space Telescope Key Project are based on the Cepheid period-luminosity relation. An alternative basis is the tip of the red giant branch. Using
archival HST data, we calibrate the infrared Tully-Fisher relation using 14
galaxies with tip of the red giant branch measurements. 
Compared with the Key Project, a higher value
of the Hubble Constant by 10\% $\pm$ 7\% is inferred. Within the errors
the two distance scales are therefore consistent. We describe the
additional data required for a conclusive tip of the red giant branch
measurement of H$_0$. 
\end{abstract}

\section{Introduction}
The extragalactic distance scale based on the Cepheid period-luminosity (PL) relation
and secondary distance indicators, such as the Tully-Fisher relation, the supernova
standard candle \citep{gib00}, surface brightness fluctuations, and the fundamental plane
\citep{{fr01},{mo00}} has been criticized recently \citep{{str08},{str07}}
on the grounds that the PL relation may not be unique. Indeed, the finite width
of the Cepheid instability strip in the HR diagram implies that nuisance parameters
such as metallity and star formation history may play a role in determining the PL
relation. Metallicity was considered as a second parameter by \cite{fr01},
\cite{sa04}, and \cite{ma06}. \cite{rom} have reviewed the situation
and concluded that the Cepheid PL relation is not universal.

It is of interest, therefore, to see how well the distance scale can be measured
without reference to Cepheids at all. In this Letter we use the tip of the red giant branch
(TRGB) distance indicator to calibrate the Tully-Fisher relation. The TRGB is a good
standard candle because it results from the helium flash on the red giant branch, which
theory suggests is relatively immune to metallicity effects in old stellar populations.

\section{TRGB distances}
The TRGB is unquestionably the most practical 
distance indicator for nearby galaxies.  It is versatile, fast, and theoretically 
verified \citep{{sc97},{mf99},{sc02}}.  

The zero point of the TRGB magnitude, however, has been debated in several 
papers.  \cite{da90} first derived the bolometric magnitude 
of the TRGB.  The distance modulus measured by the TRGB method is then 
estimated via $(m-M)_I = I_{TRGB} - M_{bol} + BC_I$, where the bolometric 
correction ($BC_I$) and the bolometric magnitude ($M_{bol}$) are both 
dependent on the color of the TRGB stars.  Using this calibration, the 
TRGB magnitude in I-band is determined to be between $-3.95$ and $-4.1$ 
depending on the 
colors of the RGB stars found.  The absolute zero point of this calibration 
was based on the distances to Galactic globular clusters that were
measured using the RR Lyrae method zero point  based on \cite{lee}.
On the other hand, \cite{sc97} presented a theoretical 
calibration of the TRGB zero point and concluded that the empirical calibration
by Da Costa \& Armandroff was too faint by $\sim$0.1 mag, likely due to the
fact that the RGB population of Galactic globular clusters used in the 
empirical calibration were not well populated around the tip.

Most recently, \cite{ri07} explored the calibration issue and 
established a new calibration based on the assumed luminosity for the 
horizontal branch and the identification of this feature in five Local 
Group galaxies.  This calibration gives the I-band TRGB magnitude of $-4.05$
at $(V-I) = 1.6$ mag.  Furthermore, the \cite{ri07} calibration is not
linked to the Cepheid distance scale in any way and is completely independent.

The sample for this section was drawn from those galaxies within 10 Mpc
with infrared photometry, distance estimates, and 21 cm data cataloged by \cite{ahm} 
and with V \& I imaging in the Hubble Space Telescope data archive (Table 1). These
images were downloaded and photometry carried out with the DAOPHOT software
of \cite{st87}. Point spread functions (PSFs) supplied by Stetson for the H$_0$
Key Project were employed \citep{kfm} for WFPC2 data. For ACS data we
used PSFs constructed from the images themselves. The ALLSTAR program was run twice
to obtain as deep a starlist as possible. Areas of the galaxy with  strong
Population I signatures were edited out. Aperture corrections were made
and color magnitude diagrams (CMDs) were calibrated and corrected for
charge transfer effects (CTE) following Dolphin\footnote{ 
http://purcell.as.arizona.edu/wfpc2\_calib} (2000). ACS data were
calibrated as described by \cite{sir} and corrected for CTE using the standard
algorithm\footnote{http://www.stsci.edu/hst/acs/performance/cte/cte\_formula\_acs\_page.pdf}. CMDs for NGC 247, 891,
4826, 4945, and 5253 are shown in figures 1--5.

For the detection of the TRGB in our target galaxies, we used the edge-detection
method described in \cite{smf}, and the results are
recorded in Table 2.  Reddening values in the table are those of \cite{sch}.
Literature values of TRGB distance moduli are from \cite{k03} and \cite{k05}.
We have preferred our value of TRGB, I = 24.10 to that of \cite{da06},
whose measurement of i$^\prime$ = 24.5 $\pm$ 0.1 can be transformed to I = 24.03,
using V--I = 1.6 and the formulae of \cite{aa02}.
Six galaxies in our sample have Cepheid distance moduli, and these are given
in the last columns of Table 2.

\section{Magnitudes and velocity widths}
The principal sources of infrared and 21 cm data are \cite{ahm} and 
\cite{sa00}. For other
galaxies we used isophotal magnitudes from the 2MASS Large Galaxy Atlas,
transforming them with H$^c_{-0.5}$ -- H$_{mk20fe}$ = 0.27 $\pm$ 0.03 mag, based
on 123 galaxies in common. For NGC 4945 and 5102 we used 20\% velocity widths from HIPASS \citep{mey} and \cite{tul} respectively, correcting them for cosmology (1+z) and inclination.
Following \cite{sa00}, we omitted the 3$^\circ$ additive term in the inclination
adopted by \cite{ahm} and \cite{tul}. Column (5) of Table 2 is the 
corrected infrared magnitude; column (6) is the corrected velocity width.

\section{Calibration of the Tully-Fisher relation}

The Tully-Fisher relation for galaxies with TRGB distances is shown in
Figure 6. The ordinate is the absolute H magnitude corrected for internal
extinction following \cite{sa00}.
 To correspond in range of velocity width to that of the cluster
galaxies to which the calibration will be applied \citep{ahm86}, we ignore galaxies
with $\Delta V(0)~<$ 200 km s$^{-1}$. The straight line in Figure 6 is
the calibration by \cite{sa00} using Cepheid distances. The mean difference
in distance modulus between the 14 TRGB galaxies and the 21 Cepheid galaxies
is 0.19 $\pm$ 0.13 mag. Applying their Cepheid Tully-Fisher calibration,
\cite{sa00} found H$_0$ = 67 $\pm$ 3 $\pm$ 10 km s$^{-1}$ Mpc$^{-1}$.
With our TRGB Tully-Fisher relation applied to the same cluster data,
we would obtain a 10\% higher value, 73 $\pm$ 5 km s$^{-1}$ Mpc$^{-1}$,
where our quoted uncertainty is the statistical error only.

\section{Discussion and conclusions}

\cite{sa00} obtained H$_0$ = 71 $\pm$ 4 $\pm$ 7 km s$^{-1}$ Mpc$^{-1}$
from their multiwavelength Cepheid-based Tully-Fisher calibration. The
largest term in the 7 km s$^{-1}$ Mpc$^{-1}$ systematic error is due to
the distance of the Large Magellanic Cloud. The largest term in the
absolute calibration of the TRGB (population II) distance scale is the
uncertainty in M$_{I,TRGB}$ = --4.05 $\pm$ 0.02 mag \citep{ri07},
associated with the absolute magnitude of the horizontal branch.

Our principal finding is that, within the 1.5$\sigma$ uncertainty, the mean
difference of the distance moduli derived from Cepheids and from the TRGB
magnitude for our sample of 14 galaxies is consistent with zero.

In addition, we conclude that the further steps to a more accurate Cepheid-independent
value of H$_0$ are (i) a larger sample of TRGB distances to galaxies
which calibrate secondary distance indicators, (ii) multiwavelength
photometry of these galaxies, and (iii) TRGB calibration of type Ia
supernovae \citep{tsr08}, surface brightness fluctuations, and the
fundamental plane.

\acknowledgements
This work is based on archival observations made with the Hubble Space Telescope, 
which is operated by the Space Telescope Science Institute under a contract with NASA. 
This work makes use of 2MASS data products, a joint project of the University of Massachusetts and IPAC/Caltech, funded by NASA and NSF. 
In addition to DAOPHOT, this research has made use of IRAF, which is distributed by NOAO. NOAO is
operated by AURA under a cooperative agreement with NSF.

\clearpage

\begin{deluxetable} {lll}
\tablewidth{0pt}
\tablecaption{HST datasets}\label{tab1}
\tablehead{\colhead{Galaxy} & \colhead{Archive dataset}& \colhead{Filter}}
\startdata
NGC 247 & j9ra78kqq, ksq, kuq, kwq & F606W, F814W\\
NGC 891 & j8eo01e9q, edq, ehq, eyq, f3q, f7q, g0q& F606W \\
        & j8eo02ofq, owq, p1q, p5q, q0q, q5q, q9q& F814W \\
NGC 4826& j9ov16uaq, ubq, udq, ufq & F606W, F814W\\
NGC 4945& u6ep1101r, 1102r, 1103r \ldots 1109r,  \\
        & 110ar, 110br, 110cr & F606W, F814W\\
NGC 5253& j9k501dbq, deq, dmq, doq & F555W, F814W\\
\enddata
\end{deluxetable}

\begin{deluxetable} {lrrrrclcl}
\tablecaption{Galaxies calibrating the Tully Fisher relation}\label{tab2}
\tablewidth{0pt}
\tablehead{
\colhead{Galaxy} & \colhead{I$_{TRGB}$} & \colhead{A$_I$} & \colhead{(m-M)$_0$} & \colhead{H$^c_{-0.5}$} & \colhead{$\Delta V_{20}$(0)} & \colhead{Ref.} & \colhead{Ceph. mod.} & \colhead{Ref.}  }
\startdata
   NGC 7793 &23.95&0.22&27.78&7.89&255&5 \nl
   NGC 224  &20.53&0.15&24.37&0.91&555&1& 24.44&a\nl
   NGC 247  &24.10&0.03&28.12&7.69&233&2& 27.80&b\nl
   NGC 253  &23.97&0.19&27.83&4.74&443&5 \nl
   NGC 598  &20.91&0.08&24.71&4.38&249&1& 24.64&a \nl
   NGC 891  &25.90&0.13&29.82&6.84&483&2 \nl
   NGC 3031 &23.91&0.16&27.70&4.38&524&1& 27.80&a \nl
   NGC 3351 &25.92&0.05&29.92&7.45&385&1,6&30.01&a \nl
   NGC 3621 &25.38&0.16&29.26&7.40&316&1& 29.13&a \nl
   NGC 4244 &     &    &28.26&8.75&221&3 \nl
   NGC 4826 &24.64&0.08&28.61&6.10&376&2 \nl
   NGC 4945 &23.95&0.10&27.90&5.16&382&2,4  \nl
   NGC 5102 &     &    &27.66&7.57&235&3,4 \nl
   NGC 5253 &23.82&0.11&27.76&8.96&103&2& 27.61&a \nl
   IC 5052   &24.84&0.10&28.80&10.24&211&7\nl
\enddata
\tablenotetext{1} {\cite{ri07}}
\tablenotetext{2} {this paper}
\tablenotetext{3} {\cite{k05}}
\tablenotetext{4} {http://irsa.ipac.caltech/applications/2MASS/LGA/}
\tablenotetext{5} {\cite{k03}}
\tablenotetext{6} {\cite{sa00}}
\tablenotetext{7} {\cite{dsj}}
\tablenotetext{a} {\cite{fe01}}
\tablenotetext{b} {\cite{gv08}}

\end{deluxetable}

\clearpage

\begin{figure}[h]
\plotone{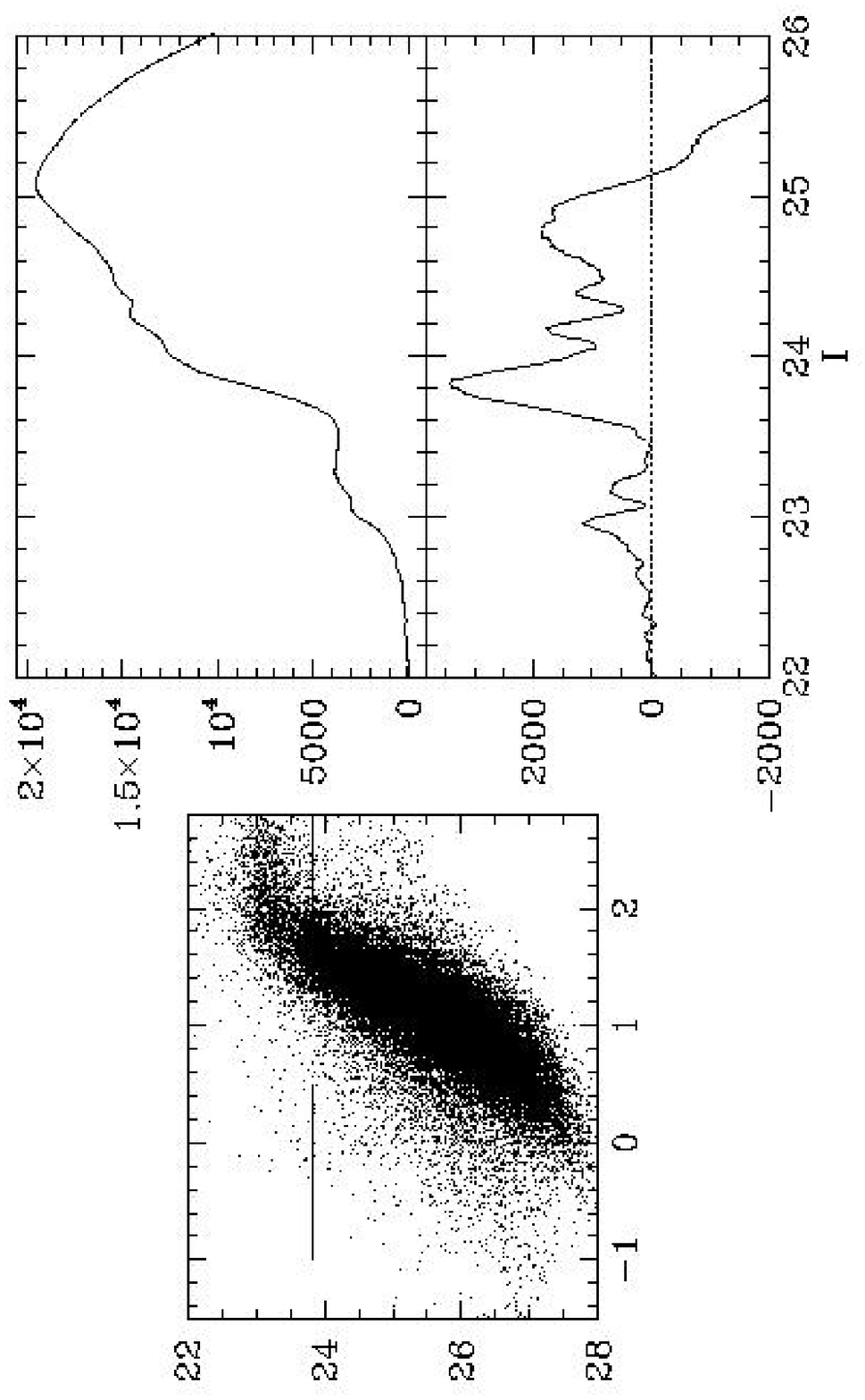}
\figcaption{(Left) the color magnitude diagram of NGC 5253 with the TRGB
marked. (Right, upper) the RGB luminosity function of NGC 5253;
(lower) the peak in the filtered indicator of the TRGB, I = 23.82 mag.}
\end{figure}

\begin{figure}[h]
\plotone{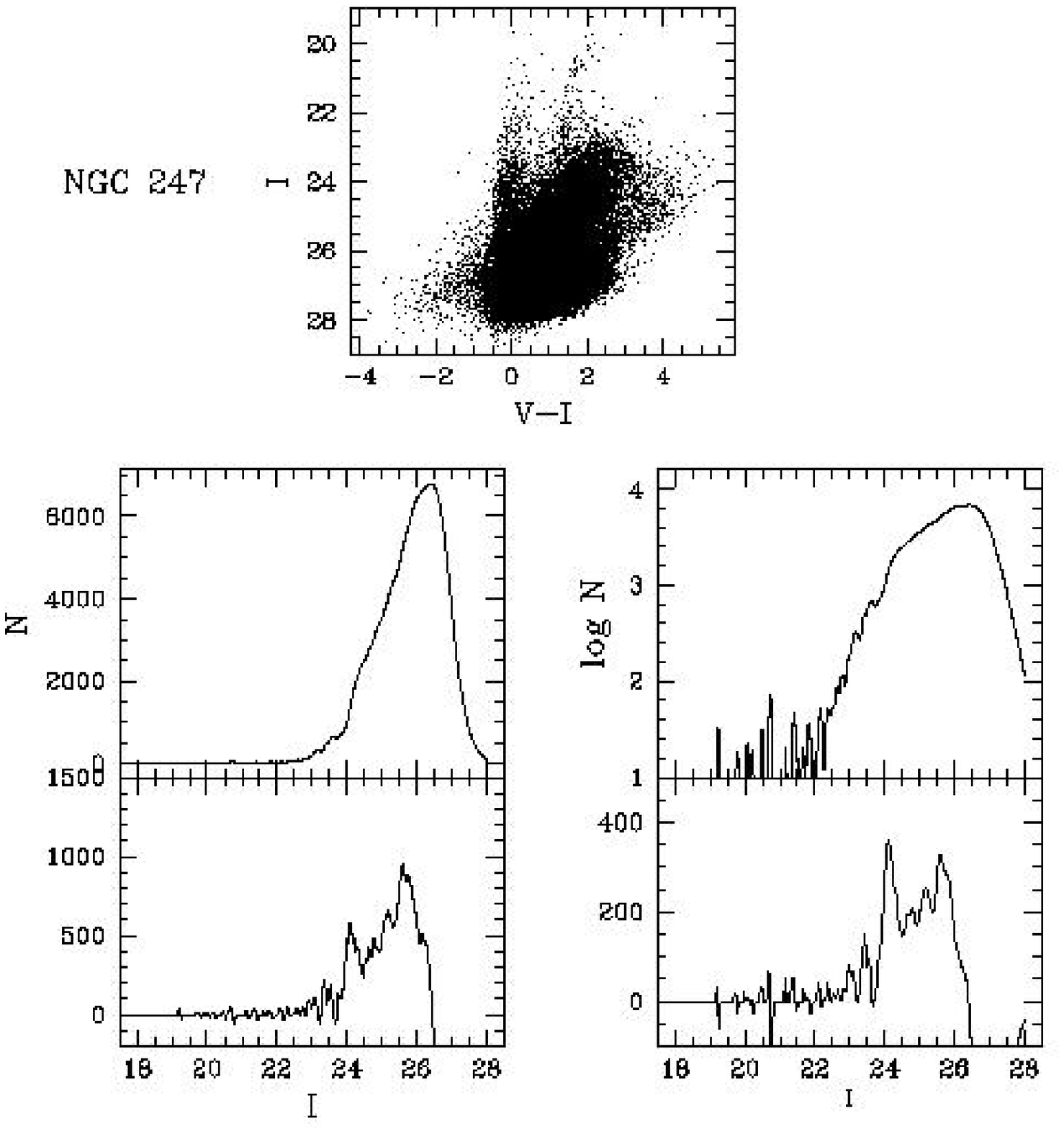}
\figcaption{(Top) the CMD of NGC 247. (Left, upper) the RGB luminosity 
function; (right, upper) the log luminosity function; (lower)
the peak in the filtered indicator of the TRGB, I = 24.10 mag.}
\end{figure}

\begin{figure}[h]
\plotone{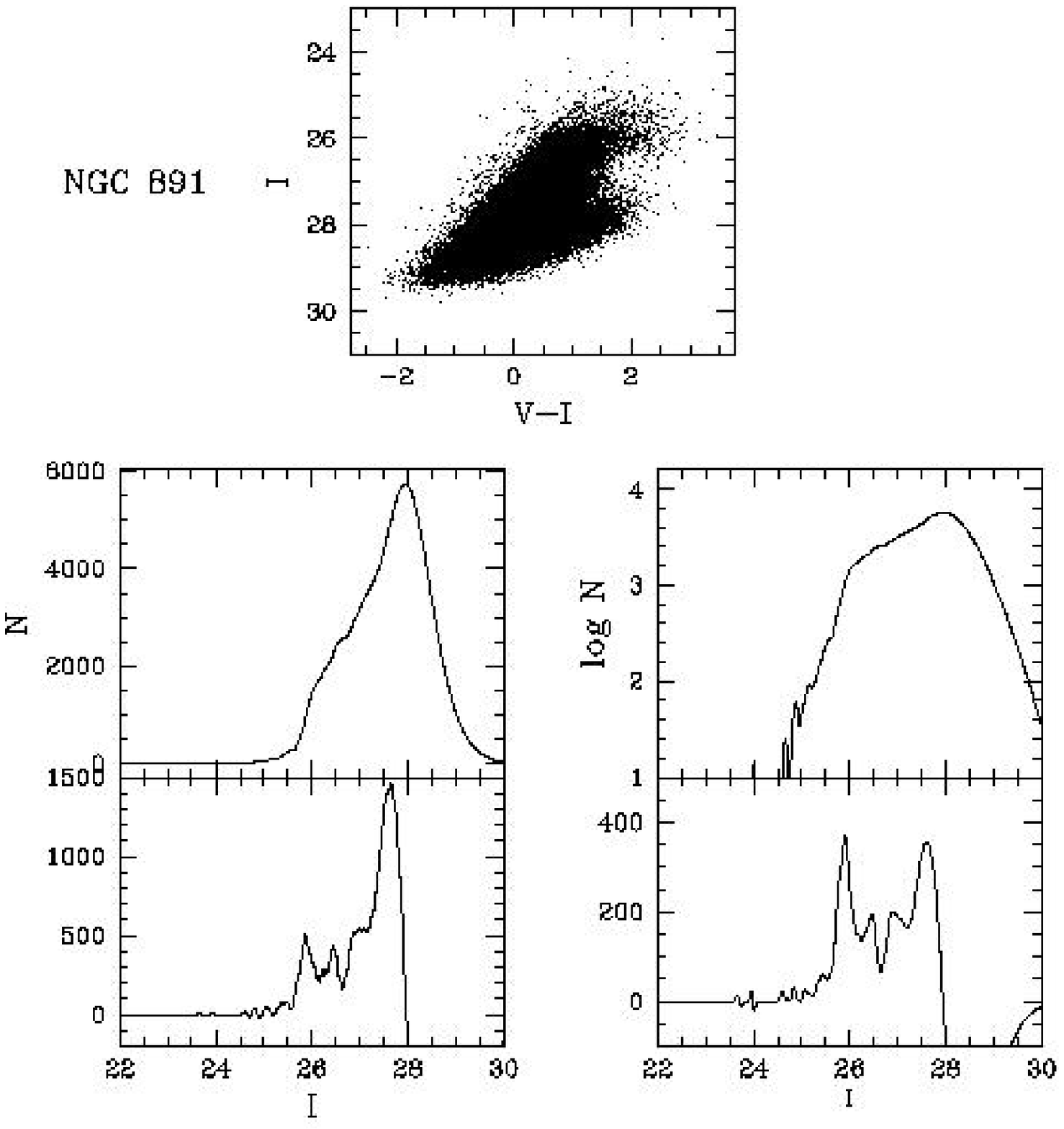}
\figcaption{(Top) the CMD of NGC 891; the other panels follow
Figure 2 with I = 25.90.}
\end{figure}

\begin{figure}[h]
\plottwo{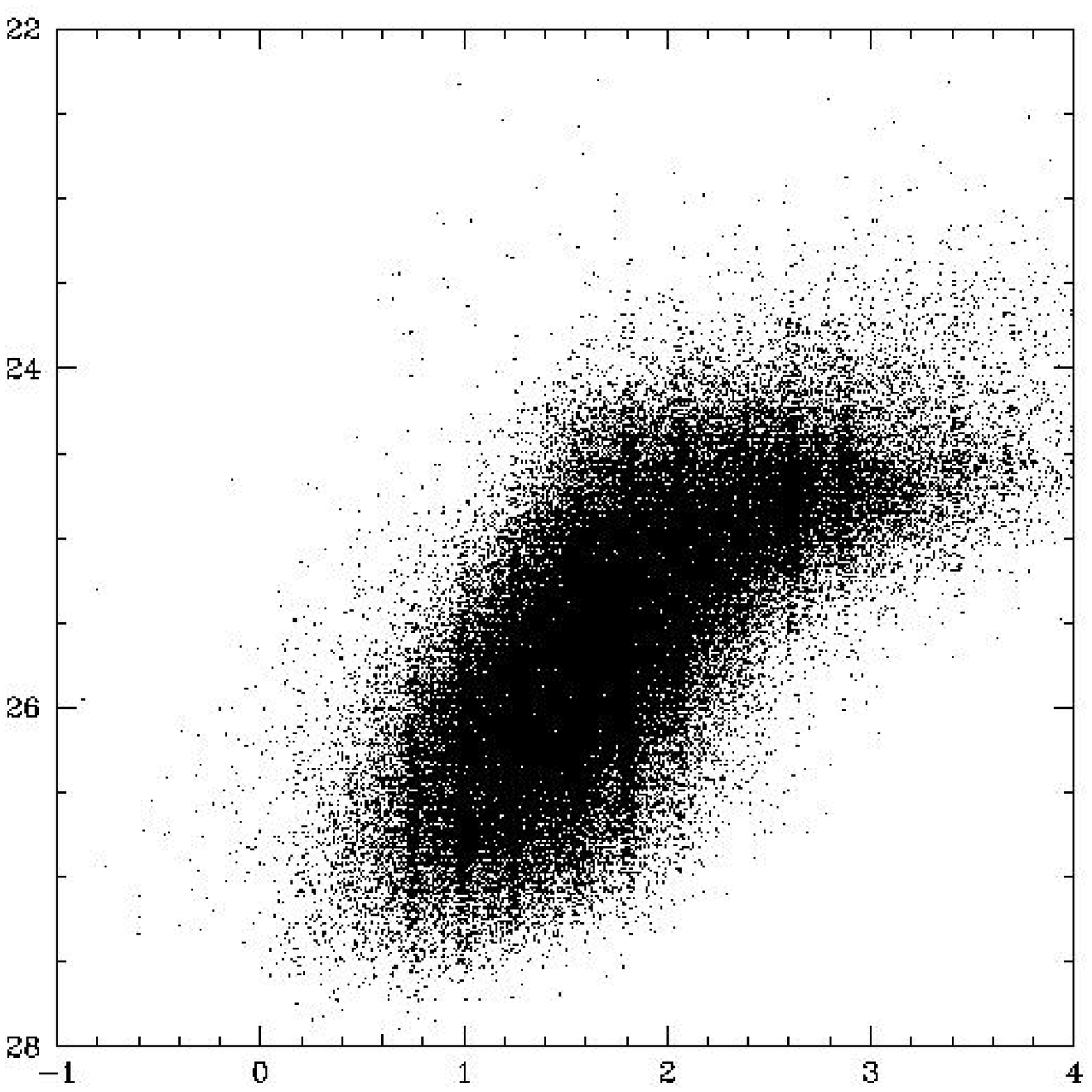}{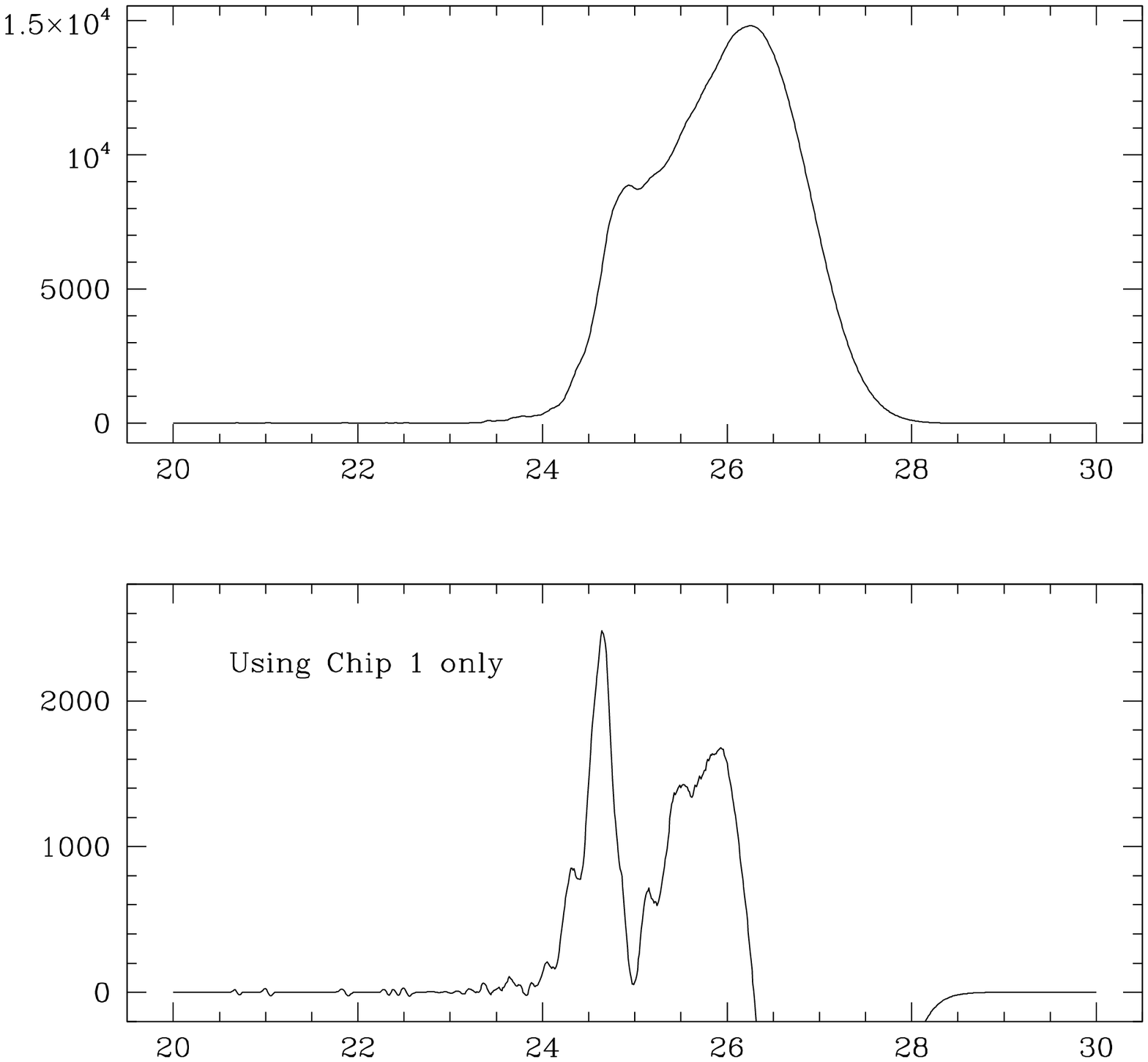}
\figcaption{(Left) the CMD of NGC 4826. (Right, upper)
the luminosity function; (lower the peak in the filtered indicator
of the TRGB, I = 24.64.}
\end{figure}

\begin{figure}[h]
\plottwo{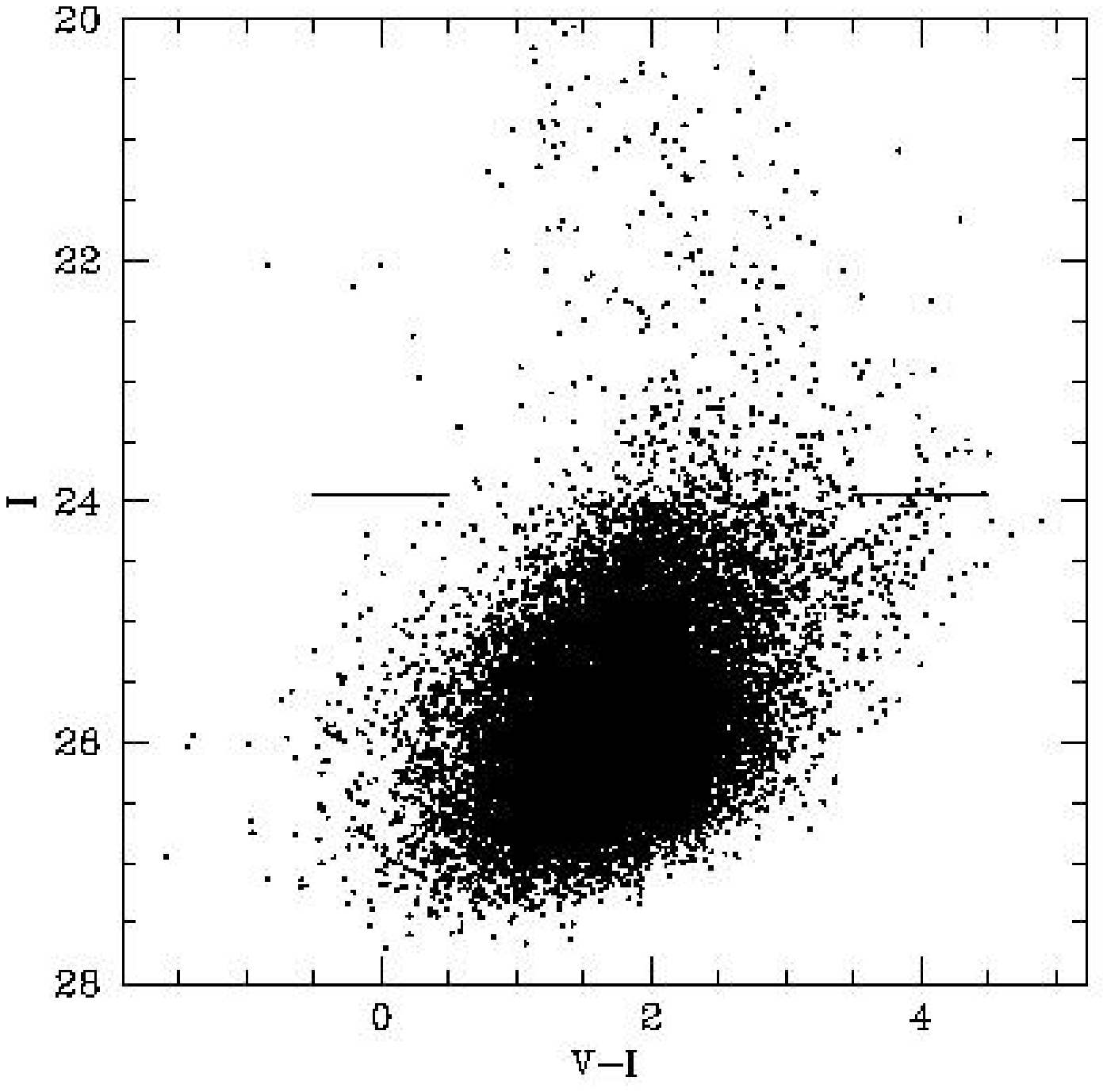}{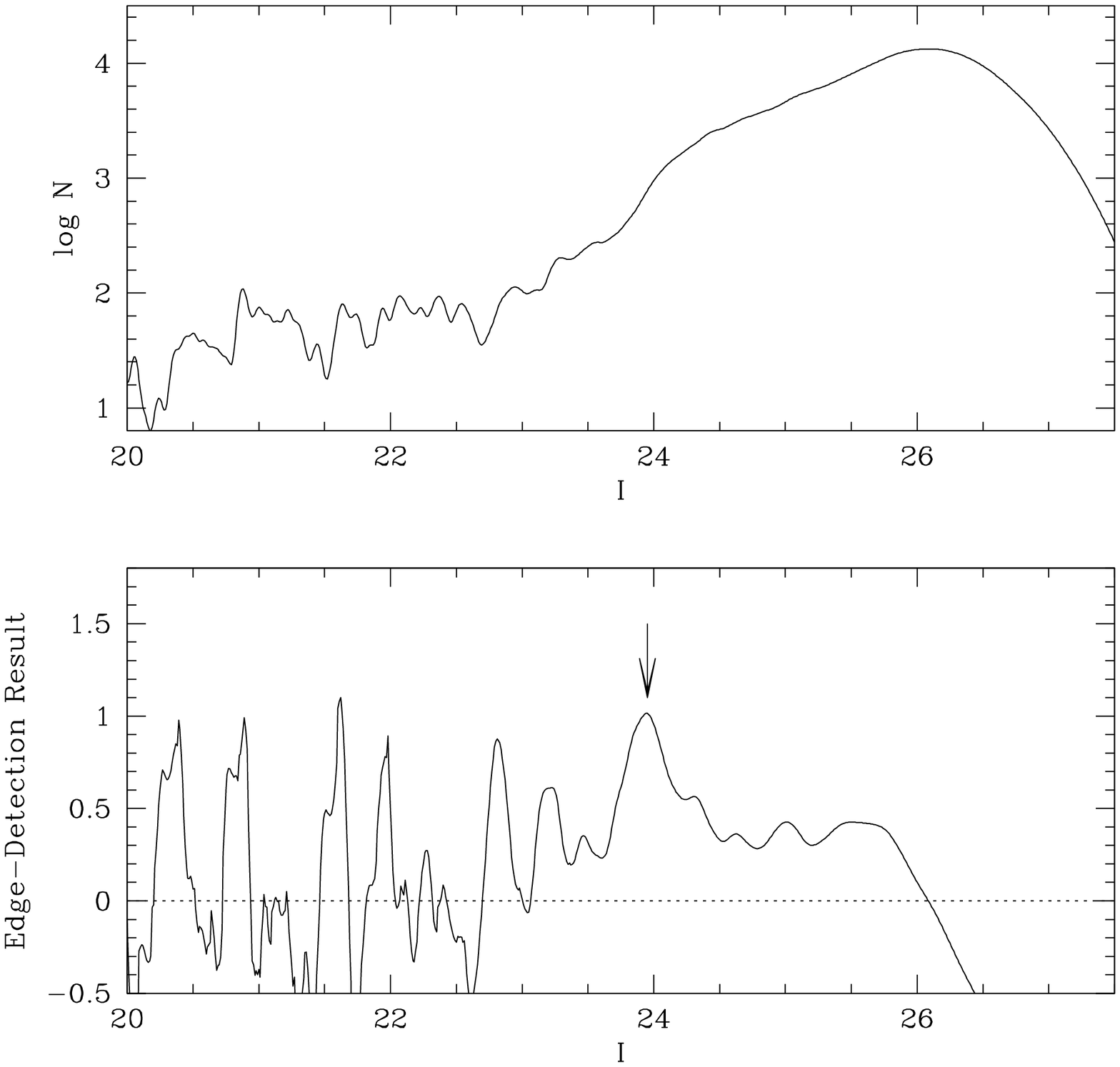}
\figcaption{(Left) the CMD of NGC 4945; the other panels follow
Figure 4 with I = 23.95.}
\end{figure}

\begin{figure}[h]
\centering{\includegraphics[angle=-90,width=\textwidth]{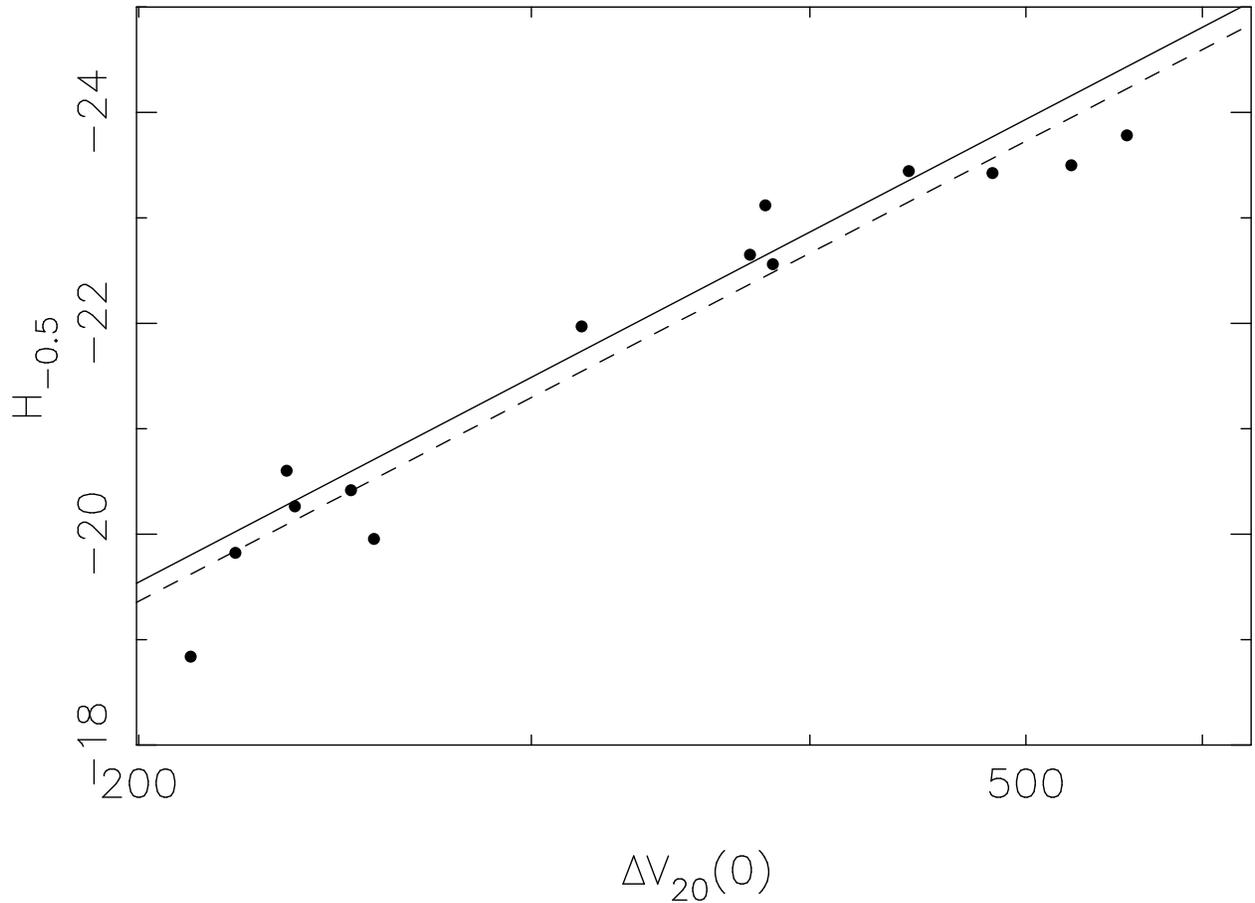}}
\figcaption{Tully-Fisher (TF) relation from the data in Table 2.
The straight line is equation 10 of \cite{sa00}
and represents the Cepheid infrared TF calibration.
The dashed line is the least-squares regression to the data points.}
\end{figure}

\end{document}